\documentclass[letterpaper]{article} % DO NOT CHANGE THIS
\usepackage{aaai25}  % DO NOT CHANGE THIS
\usepackage{times}  % DO NOT CHANGE THIS
\usepackage{helvet}  % DO NOT CHANGE THIS
\usepackage{courier}  % DO NOT CHANGE THIS
\usepackage[hyphens]{url}  % DO NOT CHANGE THIS
\usepackage{graphicx} % DO NOT CHANGE THIS
\usepackage{natbib}  % DO NOT CHANGE THIS AND DO NOT ADD ANY OPTIONS TO IT
\usepackage{caption} % DO NOT CHANGE THIS AND DO NOT ADD ANY OPTIONS TO IT
\usepackage{algorithm}
\usepackage{algorithmic}

\usepackage{amsmath,amsthm,amssymb,mathrsfs}
\usepackage{multirow}

\usepackage{newfloat}
\usepackage{listings}
\DeclareCaptionStyle{ruled}{labelfont=normalfont,labelsep=colon,strut=off} % DO NOT CHANGE THIS
\floatstyle{ruled}
\newfloat{listing}{tb}{lst}{}
\floatname{listing}{Listing}
\title{DOGR: Leveraging Document-Oriented Contrastive Learning in Generative Retrieval}
\author{
    Penghao Lu{}\textsuperscript{},
    Xin Dong\textsuperscript{}\thanks{Corresponding author.},
    Yuansheng Zhou\textsuperscript{},
    Lei Cheng\textsuperscript{},
    Chuan Yuan\textsuperscript{},
    Linjian Mo\textsuperscript{}
}
\affiliations{
    \textsuperscript{}Ant Group\\
    \{lupenghao.lph,zhaoxin.dx,zhouyuansheng.zys,lei.chenglei,yuanzheng.xy,linyi01\}@antgroup.com
}

\usepackage{bibentry}
\begin{document}

\maketitle

\begin{abstract}
Generative retrieval constitutes an innovative approach in information retrieval, leveraging generative language models (LM) to generate a ranked list of document identifiers (docid) for a given query. 
It simplifies the retrieval pipeline by replacing the large external index with model parameters. However, existing works merely learned the relationship between queries and document identifiers, which is unable to directly represent the relevance between queries and documents. 
% (ii) The mismatch between the learning objectives of identifier generation task and the document ranking objectives creates a learning gap.
To address the above problem, we propose a novel and general generative retrieval framework, namely \textit{Leveraging \textbf{D}ocument-\textbf{O}riented Contrastive Learning in \textbf{G}enerative \textbf{R}etrieval} (DOGR), which leverages contrastive learning to improve generative retrieval tasks. It adopts a two-stage learning strategy that captures the relationship between queries and documents comprehensively through direct interactions. 
% combines the generation probability of identifiers with semantic representation, jointly modeling the relevance between queries and documents from different perspective. 
Furthermore, negative sampling methods and corresponding contrastive learning objectives are implemented to enhance the learning of semantic representations, thereby promoting a thorough comprehension of the relationship between queries and documents. 
% Furthermore, contrasting learning objectives from varying views are implemented to bridge the learning gap between the generation and ranking objectives, as well as to address the document ranking issue when encountering identifier conflicts.
Experimental results demonstrate that DOGR achieves state-of-the-art performance compared to existing generative retrieval methods on two public benchmark datasets. Further experiments have shown that our framework is generally effective for common identifier construction techniques. 
\end{abstract}


\section{Introduction}

Information retrieval aims to satisfy user queries by retrieving documents. With the rapid increase in the volume of information, the importance of retrieval technology is becoming gradually prominent.
% its importance is becoming increasingly prominent as the amount of information increases dramatically. 
In recent years, a novel retrieval method known as generative retrieval has emerged. Unlike traditional retrieval methods (sparse retrieval and dense retrieval), generative retrieval assigns an identifier to each document and generates relevant document identifiers end-to-end through language models. In addition, generative retrieval can better utilize the capabilities of the large language models(LLMs), especially the knowledge learned during the pre-training phase. 

Most generative retrieval methods \cite{GENRE, DSI, NCI} first define an identifier for each document and then treated document retrieval as a standard sequence-to-sequence task. Specifically, this task learns the generation of identifiers through the interaction between queries and identifiers in an end-to-end manner. 
However, the identifier length is often much smaller than documents, identifiers may experience a certain degree of information loss compared to the complete document.
Existing generative retrieval methods mainly concentrate on
learning the relationship between the queries and the identifiers, thus they are unable to capture the complete semantic relationships between queries and documents.

There are some studies introducing additional optimization objectives to comprehensively characterize the relationship between queries and documents.
% or utilized document identifiers as a bridge.
A previous approach \cite{LTRGR} proposed a document-oriented margin-based loss based on generation probability to optimize the document ranking.
% to reduce the learning gap between the identifier and document. 
Another work \cite{GLEN} modeled query-document relevance by separately learning the generation probabilities of the query and document to the target identifier.
However, these work only focused on the intermediate target of learning identifiers and lack the interaction between query and documents.
This could potentially affect the retrieval performance.
Since the model can only learn the nuanced interactions through the identifier, it is insufficient to model the relevance between queries and documents by solely relying on the generation probability of identifiers.

To address the above problems, we introduce a novel and general framework, namely \textit{Leveraging \textbf{D}ocument-\textbf{O}riented Contrastive Learning in \textbf{G}enerative \textbf{R}etrieval} (DOGR), which captures the relationship between queries and documents comprehensively through their direct interactions.
DOGR proposed a two-stage learning strategy consisting of an identifier generation stage and a document ranking stage.
First, in the identifier generation stage, we train an encoder-decoder architecture model based on the generation loss to learn the connections between the document identifier and the query.
Next, in the document ranking stage, we fine-tune the model with contrastive learning objectives to directly learn the correlation between queries and documents. 
We enhance document representation through the elaborate design of two negative sampling methods. 
Specifically, DOGR explicitly leverages \textit{prefix-oriented negative sampling} to learn semantic representations of documents with the same identifier prefix and \textit{retrieval-augmented negative sampling} to comprehend the representation of documents with distinct identifier prefixes.
Drawing on the above mentioned negative samples, two corresponding contrastive losses are utilized to enhance the semantic similarity between query and its positive document, aiming to achieve analogous representations for both by increasing direct interaction.
During the inference phase, we not only consider the generation probability of document identifiers, but also leverage the semantic score to jointly model the relevance between the query and documents. 
By means of ranking the documents corresponding to the generated identifiers, the performance of generative retrieval is further enhanced.

In summary, our main contributions are as follows:
\begin{itemize}
    \item We propose \textbf{DOGR}, a novel and general framework which adopts contrastive learning to enhance generative retrieval. It incorporates a two-stage learning strategy that jointly models the relevance of query-document by combining the generation probability of identifiers with document semantics.
    \item According to the characteristics of generative retrieval, we introduced two negative sampling methods and corresponding contrastive learning objectives to help the model fully capture the correlation between query and documents through direct interaction.
    % 帮助模型通过query和document的直接交互来刻画它们之间的关系
    % to address the ranking problem of documents under identifier conflicts, as well as to optimize the document ranking objective through contrastive learning.
    \item Comprehensive experiments show that \textbf{DOGR} outperforms existing state-of-the-art generative retrieval methods on two benchmark datasets. Further experiments have demonstrated that our method is effective for several common identifier construction approaches.
\end{itemize}
\begin{figure*}[!htb]
\centering
\includegraphics[height=11.08cm,width=17.5cm]
{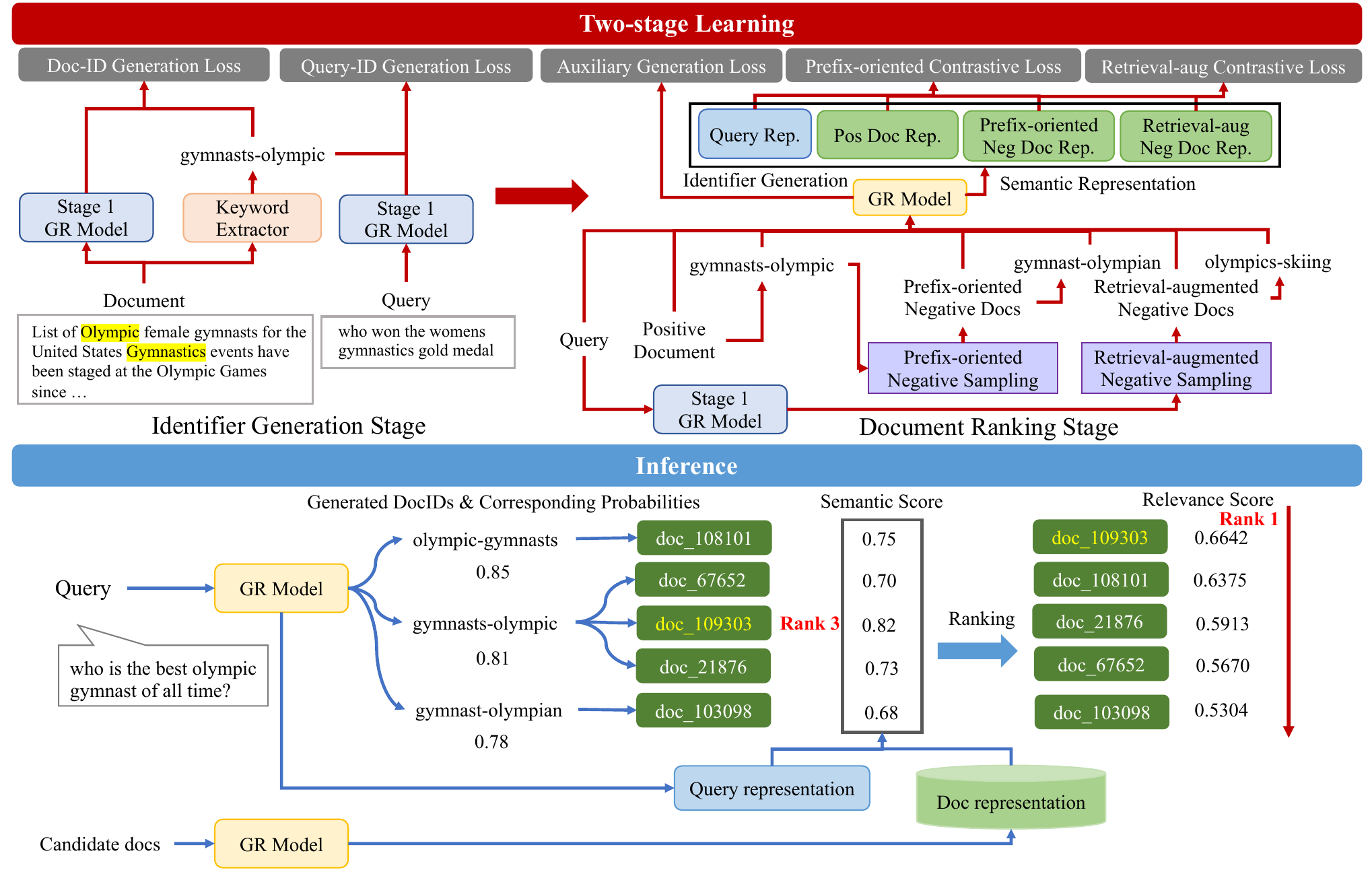}
\caption{An overview of the framework of DOGR. The training phase can be divided into two stages. The first is the identifier generation stage, which is used to learn the relationship between text and document identifiers. The second is the document ranking stage, which employs contrastive learning to learn semantic representation for document ranking. In the inference phase, DOGR first generates identifiers for the given query, then the documents are ranked according to the relevance scores derived from generation probabilities of identifiers and the semantic scores.}
\label{framework}
\end{figure*}

\section{Related Work}
\subsection{Document Retrieval}
Traditional document retrieval methods mainly include sparse retrieval and dense retrieval.
Sparse retrieval (classic methods such as BM25 \cite{BM25} and TF-IDF \cite{TFIDF}) has been widely used in industry due to its high efficiency.
% , it usually calculates weights by establishing an inverted index and using statistical information such as term frequency. 
However, sparse retrieval relies on lexical overlap and finds it challenging to match semantically related but literally different documents.
% , while there is also a considerable gap in the calculation of term weights from perfection.
Dense retrieval involves representing queries and documents as dense vectors\cite{dense1}, and calculating the similarity between the query and the document using inner product or cosine distance. 
Benefiting from the continuous development of pre-trained language models, the performance of dense retrieval is also constantly improving. Contrastive learning\cite{contrastive1, contrastive2} and hard negative sampling techniques \cite{ANCE,DPR,SentenceT5,GTR-base} have become another important driving force in the development of dense retrieval.
% 对比学习和困难样本挖掘技术成为稠密检索发展的另一重要助力。
Therefore, it has gradually become another mainstream retrieval method in industrial implementation. Although dense retrieval has achieved great success, it suffers from issues such as embedding space bottleneck and a lack of fine-grained interactions between embeddings. 
% While generative retrieval still lags behind state-of-the-art dense retrieval methods, it presents a new approach with significant exploration potential.
% 尽管稠密检索获得了巨大成功，稠密检索 suffer from issues such as embedding space bottleneck and lack of fine-grained interactions between embeddings. 虽然生成式检索仍落后于最先进的稠密检索方法，但它是一种新的检索方式存在巨大的探索空间。
% However, dense retrieval is difficult to optimize end-to-end due to its expensive computational cost.
\subsection{Generative Retrieval}
Different from traditional document retrieval methods, generative retrieval uses only the language model to directly generate the identifiers of relevant documents for a given query. GENRE \cite{GENRE} is the first step in performing information retrieval tasks using an autoregressive language model. It implemented constrained beam search based on a prefix tree to generate entity titles within a given candidate set. 
Since document identifiers largely determines the effectiveness of generative retrieval, many studies have explored the construction methods of identifiers. 
In general, identifiers can be categorized into lexical and numerical types. Lexical identifiers allow for explicit expression including n-grams\cite{SEAL,NOVO}, titles\cite{GENRE,MINDER}, etc, while numerical identifiers\cite{DSI,NCI,Genret} mainly capture similarities and differences among documents through clustering methods. 
However, these works focused on the learning process of mapping queries to identifiers, which cannot directly represent the relevance between queries and documents.

Recently, there have been studies that introduce additional optimization objectives to help characterize the relationship between queries and documents.
LTRGR \cite{LTRGR} proposed a margin-based ranking loss based on generation probability from the document level to improve document ranking.
GLEN \cite{GLEN} utilized identifiers as a bridge to model query-document relevance by separately learning the generation probabilities of both the query and the document in relation to the target identifier.
Although they open up new ideas for generative retrieval, they still only utilized the generation probability of document identifiers, which has a lack of direct interaction between the query and the document.
To address the aforementioned issues, we propose a two-stage learning strategy to learn the generation probabilities at the identifier level and the semantic representation at the document level.  
Through direct interaction between queries and documents, we enhance generative retrieval through contrastive learning objectives based on negative sampling methods.

\section{The Proposed Scheme}
In this work, we propose a two-stage learning strategy that enhances query representations to better align with document representations for generative retrieval.
Figure~\ref{framework} shows an illustration of the core procedure of DOGR, including an identifier-level generation stage and a document-level ranking stage.
First, in the \textit{Identifier Generation Stage}, DOGR trains an encoder-decoder architecture model to learn the connections between the keyword-based document identifier with query. 

Then, in the \textit{Document Ranking Stage}, DOGR fine-tunes the trained model by introducing contrastive learning to depict the relationship between the query and the complete document.
We improve document representation by elaborately designing two negative sampling methods.
In particular, we use prefix-oriented negative sampling to capture the semantic representations of documents sharing the same identifier prefix. On the other hand, we propose retrieval-augmented negative sampling to better understand the representation of documents with different identifier prefixes.
Informed by the aforementioned negative samples, two related contrastive losses are applied to maximize the semantic score between the query and the positive document. This approach strives to attain comparable representations for both.

In the inference phase, both the generation probability of document identifiers and the semantic score are taking into account to collaboratively model the relevance between query and documents. 
Through ranking the documents corresponding to the generated identifiers, the performance of generative retrieval is further enhanced. 

\subsection{Identifier Generation Stage}
\subsubsection{Keyword-Based Lexical Identifier.}
Existing generative retrieval methods require constructing identifier for each document.
Since keywords are important carriers of document information, we extract keywords as the document identifier.
Specifically, for each document $d$, pre-trained KeyBert \cite{KeyBert} is used to extract the top keywords and sort the keywords in descending order of importance. After encoding the keyword sequence with the tokenizer, the first $n$ tokens are selected as the document identifier $z^d = \{z_0^d, z_1^d, \cdots, z_{n-1}^d \}$. As a result, each document $d$ has its corresponding identifier $z^d$. While a document identifier $z$ can correspond to a collection of documents $\{d_i | z^{d_i} = z\}$, since similar documents may have the same identifier.
In addition to keyword-based identifier, our framework is applicable to common identifier types including both lexical and numerical type, and its generalization will be validated in the experiment section.

\subsubsection{Pointwise Generation Task.} 
Similar to the previous identifier generation tasks, the training objective in this stage is a standard sequence-to-sequence task. This task aims at learning the relationship between query and document identifiers. 

% 对于训练集中的样本(q,d)，我们采用来最大化 xx的loglikehoods
For each sample $(q, d)$ in training data, we adopt sequence-to-sequence cross-entropy loss with teacher forcing to maximize the log-likelihoods of $q \rightarrow z^d$.
It is far from sufficient to simply model the generation task of query to target document identifier. This is because the query cannot represent the entirety of the document and it cannot assist the model in fully learning the relationship between the document and the target identifier, which may lead to poor robustness. Thus we utilize the same loss function to maximize the loglikelihoods of $d \rightarrow z^d$. Different from the query input, document segments are randomly sampled as input to learn the relationship between document and its identifier more comprehensively. 
It was observed that training the document and query tasks together in a multi-task setting leads to better performance in generating identifiers than training these tasks separately. Accordingly, the generation loss is formulated as follows,
\begin{equation}
    L_{1} = -\sum_{t=1}^{n}\log{P(z^d_t|q,z^d_{<t})} -\sum_{t=1}^{n}\log{P(z^d_t|d,z^d_{<t})},
\end{equation}
where $z^{d}_t$ is the token at $t$-th step of $z^d$. 
\subsection{Document Ranking Stage}
Since the identifier generation task mainly focuses on learning the relationship between queries and identifiers, which has the limited ability to comprehensively understand the relationship between queries and documents.
Therefore, the purpose of the document ranking stage is to learn the semantic representation of document through its direct interaction with the query. In light of the characteristics of generative retrieval, two negative sampling strategies and corresponding contrastive learning objectives are implemented to optimize document representation.
\subsubsection{Prefix-oriented Negative Sampling.}
We introduce contrastive loss to help the model to represent queries as close to positive documents $d^+$ and far from negative documents $d^-$.
Then, we propose prefix-oriented negative sampling to sample negative documents with the same identifier prefix within the batch.
Specifically, we first select documents with the same first $n$ tokens as $z^{d^+}$ (excluding $d^{+}$ itself).
Next, documents that have the same first $n-1$ tokens as $d^{+}$ are fetched. We repeatedly reduce the prefix length until $|\mathcal{N}_z|$ negative samples are obtained.
Since identifiers with the same prefix tend to have similar document information,
% is a commonly used negative sampling strategy in generative retrieval \cite{GLEN}, 
this negative sampling method can help the model better distinguish difficult samples. 
% Meanwhile, it helps to solve the document ranking problem when encountering identifier conflicts.
Based on the semantic representation, we define the semantic score between the query and document as follows, 
\begin{equation}
    s(q, d) = h_q \cdot h_d^T,
\end{equation}
% \begin{equation}
%     h_x = {\rm AvgPool}({\rm Enc}(x))
% \end{equation}
where $h_x \in \mathbb{R}^m$ is the average pooling of the hidden representations at each position of the encoder for a given input $x$. According to the relationship between the positive and negative samples mentioned above, the 
prefix-oriented contrastive learning loss function can be formulated:
\begin{small}
\begin{equation}
\nonumber
    L_{c_i} = -\log{\frac{\exp(s(q,d^+)/\tau)}{\exp(s(q,d^+)/\tau) + \sum_{d^- \in \mathcal{N}_z}\exp(s(q,d^-)/\tau)}},
\end{equation}
\end{small}
where $\tau$ is the temperature and $\mathcal{N}_z$ is the set of negative documents obtained via prefix-oriented negative sampling.
\subsubsection{Retrieval-augmented Negative Sampling.} 
% \
% \newline
% However, only negative sampling for the same prefix is far from enough, because the document identifiers generated via beam search during inference stage are diverse and do not necessarily have the same prefix. 
% The ability to rank documents with different identifier prefixes needs further improvement. 
% Therefore, we introduce retrieval-augmented negative sampling. 
Then, we introduce retrieval-augmented negative sampling to sample negative documents with distinct identifier prefixes.
To be specific, it samples negative documents by generating high-ranking identifiers for the given query. 
For each query $q$ during training, the top $k$ document identifiers $\{z_1, \cdots, z_k\}$ are retrieved based on the model trained in the first stage. The document collection corresponding to these identifiers is $\bigcup_{j=1}^{k}\{d_i | z^{d_i} = z_j\}$. 
% 在这些文档中我们将除去与正样本文档的标识符一样的文档集合，因为它们已经在prefix-oriented negative sampling采样过了。
In these documents, the documents that have the same identifier as the positive document are removed, as they have already been sampled in the prefix-oriented negative sampling.
 Then the negative documents $\mathcal{N}_q$ are sampled from the remain documents, i.e. $\bigcup_{j=1}^{k}\{d_i | z^{d_i} = z_j\} / \{d_i | z^{d^+} = z_j\}$ , where $|\mathcal{N}_q| < k$. Thanks to the negative samples provided by the first stage learning, a retrieval-augmented contrastive learning loss function is formulated as follows,
\begin{small}
\begin{equation}
\nonumber
    L_{c_q} = -\log{\frac{\exp(s(q,d^+)/\tau)}{\exp(s(q,d^+)/\tau) + \sum_{d^- \in \mathcal{N}_q}\exp(s(q,d^-)/\tau)}}.
\end{equation}
\end{small}
\\
\textbf{Training.}
%The generation loss in the first stage is still retained. 
We also incorporate the generation loss into the document ranking stage.
It ensures that the model's ability to generate document identifiers does not deteriorate due to the introduction of ranking learning objectives. 
Moreover, we use an additional loss to learn the corresponding document identifiers for the negative documents obtained by the retrieval-augmented negative sampling. 
We add this loss to the generation loss $L_1$ as the auxiliary generation loss $L_g$, which is calculated as follows,
% \begin{equation}
% \begin{aligned}
%      L_{g} = & -\sum_{t=1}^{n}\log{P(z^d_t|q,z^d_{<t})} \\
%     & -\sum_{t=1}^{n}\log{P(z^{d^+}_t|d^+,z^{d^+}_{<t})} \\
%     & -\sum_{d^-\in N_q}\sum_{t=1}^{n}\log{P(z^{d^-}_t|d^-,z^{d^-}_{<t})}
% \end{aligned}
% \end{equation}
\begin{equation}
    L_g = L_1 + (\sum_{d^-\in \mathcal{N}_q}\sum_{t=1}^{n}-\log{P(z^{d^-}_t|d^-,z^{d^-}_{<t})}).
\end{equation}

Therefore, the final loss function is defined as,
\begin{equation}
    L = L_{c_i} + L_{c_q} + \lambda_{g} \cdot L_{g},
\end{equation}
% \begin{equation}
%     L_g = L_1 + (\sum_{d^-\in \mathcal{N}_q}\sum_{t=1}^{n}-\log{P(z^{d^-}_t|d^-,z^{d^-}_{<t})}),
% \end{equation}
where $\lambda_g$ is the hyperparameter used to control the importance of generation loss, which balances the generation objective and contrastive learning objectives.
\subsection{Inference}
During the inference phase, we integrate the identifier generation probabilities with the semantic representation to retrieve and rank documents.

First, for a given query $q$, we apply constrained beam search \cite{GENRE} to generate a ranked list of top $k$ document identifiers $\{z_1, z_2, \cdots, z_k\}$ with the corresponding generation probabilities $P(z_1 | q), P(z_2 | q), \cdots, P(z_k | q)$. 
Next, we calculate the query representation $h_q$ and document representations $\{h_d | d \in \mathcal{D}_q\}$, where $\mathcal{D}_q$ is the corresponding document collection to the top $k$ document identifiers.
Based on the generation probability of the identifier and overall semantic information, the fusion relevance score between query and document is defined as follows:
\begin{equation}
    rel(q, d) = P(z^d | q)\cdot s(q, d).
\end{equation}
Finally, we rank the obtained document collection $\mathcal{D}_q$ based on their relevance scores and the top $k$ documents with the highest relevance scores are selected as the final retrieval results for the given query. Meanwhile, the document ranking issue under identifier conflicts can be simply addressed.
\begin{table*}[!htb]
\centering
\setlength{\tabcolsep}{0.45mm}{
\begin{tabular}{l|ccc|ccc|ccc}
\hline
 & \multicolumn{3}{c|}{NQ320K(7830)} & \multicolumn{3}{c|}{Seen test(6075)} & \multicolumn{3}{c}{Unseen test(1755)} \\
Model & \multicolumn{1}{c}{R@1} & \multicolumn{1}{c}{R@10} & \multicolumn{1}{c|}{MRR@100} & \multicolumn{1}{c}{R@1} & \multicolumn{1}{c}{R@10} & \multicolumn{1}{c|}{MRR@100} & \multicolumn{1}{c}{R@1} & \multicolumn{1}{c}{R@10} & \multicolumn{1}{c}{MRR@100} \\ \hline
 Sparse \& Dense Retrieval  \\ \hline
% Most Popular & 1.368 & 2.259 & 3.020 & 2.400 & 3.936 & 5.226 & 0.395 & 2.065 & 5.424 & 0.735 & 3.603 & 9.309 \\
BM25\cite{BM25} & 29.7 & 60.3 & 40.2 & 29.1 & 59.8 & 39.5 & 32.3 & 61.9 & 42.7 \\
DocT5Query\cite{DocT5Query} & 38.0 & 69.3 & 48.9 & 35.1 & 68.3 & 46.7 & 48.5 & 72.9 & 57.0 \\
DPR\cite{DPR} & 50.2 & 77.7 & 59.9 & 50.2 & 78.7 & 60.2 & 50.0 & 74.2 & 58.8 \\ 
ANCE\cite{ANCE} & 50.2 & 78.5 & 60.2 & 49.7 & 79.2 & 60.1 & 52.0 & 75.9 & 60.5 \\ 
SentenceT5\cite{SentenceT5} & 53.6 & 83.0 & 64.1 & 53.4 & 83.9 & 63.8 & 56.5 & 79.5 & 64.9 \\ 
GTR-base\cite{GTR-base} & 56.0 & 84.4 & 66.2 & 54.4 & 84.7 & 65.3 & 61.9 & 83.2 & 69.6 \\ \hline
Generative Retrieval  \\ \hline
GENRE\cite{GENRE} & 55.2 & 67.3 & 59.9 & 69.5 & 83.7 & 75.0 & 6.0 & 10.4 & 7.8 \\
DSI\cite{DSI} & 55.2 & 67.4 & 59.6 & 69.7 & 83.6 & 74.7 & 1.3 & 7.2 & 3.5 \\
SEAL\cite{SEAL} & 59.9 & 81.2 & 67.7 & - & - & - & - & - & - \\
DSI-QG\cite{DSIQG} & 63.1 & 80.7 & 69.5 & 68.0 & 85.0 & 39.5 & 32.3 & 61.9 & 42.7 \\
NCI\cite{NCI} & 66.4 & 85.7 & 73.6 & 69.8 & 88.5 & 76.8 & 54.5 & 75.9 & 62.4 \\
GENRET\cite{Genret} & 68.1 & \underline{88.8} & \underline{75.9} & 70.2 & \underline{90.3} & 77.7 & \textbf{62.5} & \underline{83.6} & \textbf{70.4} \\
LTRGR\cite{LTRGR} & 67.5 & 86.2 & 74.8 & 70.2 & 88.7 & 77.3 & 58.1 & 77.3 & 66.2 \\
GLEN\cite{GLEN} & \underline{69.1} & 86.0 & 75.4 & \underline{72.5} & 88.9 & \underline{78.5} & 57.6 & 75.9 & 63.9 \\ \hline
\textbf{DOGR}(Ours) & \textbf{70.2} & \textbf{89.1} & \textbf{76.8} & \textbf{72.6} & \textbf{90.5} & \textbf{78.9} & \underline{61.9} & \textbf{84.3} & \underline{69.5} \\ \hline
\end{tabular}
}
\caption{Performance comparison for the proposed framework and baseline methods for NQ320k. Bolded numbers are the best performance of each column and the second best method is underlined. The number in parentheses indicates the number of queries. We refer to the results of baselines reported by \cite{GLEN}. All the numbers in the table are percentage numbers with `$\%$` omitted. Results not available are denoted as '–'.}
\label{model performance NQ320K}
\end{table*}
\section{Experiments}
\subsection{Datasets}
\textbf{Natural Questions(NQ320k)} \cite{NQ320K} contains 320k training data (relevant query-document  pairs), 100k documents, and 7,830 test queries, and is widely used in existing generative retrieval methods. We follow the same setup as previous work \cite{GLEN} and split the test set into two subsets: seen test and unseen test. The seen test consists of queries with target documents that are covered in the training data, while the documents in the unseen test are not covered.
\\
\textbf{MS MARCO passage ranking (MS MARCO)} \cite{MSMARCRO} is a large-scale benchmark dataset that includes 8.8 million passages collected from Bing search results and 1 million real-world queries, with the test set containing 6980 queries.
\subsection{Metrics}
According to the evaluation methods of previous works, we adopt the widely accepted metrics, namely Recall@1, Recall@10, and Mean Reciprocal Rank (MRR), to comprehensively assess retrieval performance. 
% Recall@k quantifies the probability of successfully retrieving the target document within the top $k$ predicted documents. MRR expresses the ordinal position of the retrieved target document within the predicted documents.
\subsection{Baselines}
% We compared with traditional document retrieval methods (including two types of sparse retrieval and four types of dense retrieval) and eight types of generative retrieval baseline models.
We compare with several types of baseline models, including traditional document retrieval methods and generative retrieval models.
\subsubsection{Traditional Document Retrieval Methods}
In traditional document retrieval methods, we consider several commonly used highlighting models, covering both sparse retrieval (BM25 \cite{BM25}, DocT5Query \cite{DocT5Query}) and dense retrieval techniques (ANCE \cite{ANCE}, DPR \cite{DPR}, SentenceT5 \cite{SentenceT5}, GTR-base \cite{GTR-base}).
\subsubsection{Generative Retrieval Methods}
Furthermore, we compare DOGR with several generative retrieval baselines including GENRE \cite{GENRE}, DSI \cite{DSI}, SEAL \cite{SEAL}, DSI-QG \cite{DSIQG}, NCI \cite{NCI}, GENRET \cite{Genret}
% , GLEN \cite{GLEN} and LTRGR \cite{LTRGR}. 
, and two methods (GLEN \cite{GLEN} and LTRGR \cite{LTRGR}) introducing additional objectives to model query-document relationship.  
For the sake of experimental fairness, we replaced the backbone of LTRGR from BART \cite{BART} to T5-base \cite{T5} for training.
\subsection{Implementation Details}
\textbf{Keyword-based Lexical Identifier}
For each document, we first extract the top 20 diverse keywords based on KeyBert using maximal marginal relevance, while the diversity parameter is set to 0.5 and n-gram is set to 1. After experimenting with different identifier token length, we set the token length of document identifiers to 8 and 12 for NQ320k and MS MARCO respectively. We will analyze in detail the impact of identifier length on retrieval performance below.
\subsubsection{Query Generation}
It has been proved that augumented queries related to the document can effectively help the model capture the relationship between query and documents. Therefore, we follow the previous work \cite{NCI} and augment the query through Document as Query and DocT5Query.

\subsubsection{Training and Inference}
To ensure a fair comparison with previous work, we use T5-base \cite{T5} as the generative language model backbone.
In the training phase, batch size is set to 256 and 32, and the model is optimized for up to 3M and 1M steps using the Adam optimizer with learning rates 5e-5 for identifier generation stage and document-level ranking stage, respectively. The number of negatives from retrieval-augmented negative sampling is set to 4 per query, while the prefix-oriented negative sampling employs in-batch negatives. In the document ranking stage, $\tau$ is set to 0.5 as the temperature parameter for contrastive learning, and $\lambda_g$ is set to 0.1 to balance the generative task and the contrastive learning task. In the inference phase, we use the beam search with constrained decoding during inference and set the beam size to 100.
Our experimental code is implemented on Python 3.8 using transformers 4.37.0, while experiments are conducted on 6 NVIDIA A100 GPUs with 80 GB of memory.
\begin{table}[!t]
\setlength{\tabcolsep}{2.0mm}{
\begin{tabular}{l|c}
\hline
 & \multicolumn{1}{c}{MS MARCO Dev(6980)} \\
Model & MRR@10 \\ 
% \multirow{4}{*}{Amazon Books} & MiceRec(K=2) & 10.384 & 16.344 & 20.781 \\
\hline
BM25 & 18.4  \\
DocT5Query & 27.2 \\ 
GTR-base\cite{GTR-base} & 34.8 \\ \hline
DSI\cite{DSI} & 3.1  \\
DSI-QG\cite{DSIQG} & 11.8 \\
NCI\cite{NCI} & 17.4  \\ 
LTRGR\cite{LTRGR} & \underline{21.6} \\
GLEN\cite{GLEN} & 20.1 \\ \hline
\textbf{DOGR} & \textbf{22.5}  \\ \hline
\end{tabular}
}
\caption{Performance comparison for the proposed framework and baseline methods for MS MARCO. Bolded numbers are the best
performance of each column and the second best method is underlined. All the numbers are percentage numbers with `$\%$` omitted.}
\label{model performance MSMARCO}
\end{table}
\subsection{Experimental Results}
\subsubsection{Main Results}
\ 
\newline
\textbf{Evaluation on NQ320k.} Table \ref{model performance NQ320K} shows the retrieval performance on NQ320k. The main observations are as follows:
\begin{itemize}
    \item DOGR surpasses traditional document retrieval methods and outperforms existing generative retrieval methods. Specifically, DOGR outperforms the best competitive generative retrieval model on the full test set of NQ320k by +1.6\%, +0.3\%, and +1.2\% on Recall@1, Recall@10 and MRR@100, respectively. 
    \item DOGR not only outperforms existing baseline models on the full and seen test sets, but also achieves competitive performance with the previous best model on the unseen test set, which shows that our method has good robustness.
    % DOGR不仅在full 和seen test set上超过了其他模型，在unseen test上也取得了和之前最佳模型具有竞争力的表现。这表明了我们的方法具有不错的鲁棒性。
    \item Our method exhibits higher performance than other methods adopting additional objectives to model query-document relationship in generative retrieval. This demonstrates the effectiveness of enhancing generative retrieval via document-oriented contrastive learning.
    % 我们的方法 exhibit higher performance than 其他在生成式检索中采用ranking的方法。这说明了通过对比学习增强生成式检索的有效性。
\end{itemize}
\textbf{Evaluation on MS MARCO}. Table \ref{model performance MSMARCO} presents the retrieval performance on MS MARCO. DOGR improves MRR@10 by 4.2\% compared to the best competitive generative retrieval method and by 22.3\% compared to BM25\cite{BM25}. Existing generative retrieval methods still struggle to memory knowledge of the document corpus solely through identifiers, making it difficult to operate effectively in large-scale corpora. In contrast, DOGR enhances generative retrieval through contrastive learning and document semantics, thus it is successfully performed in large-scale corpora.
\subsubsection{In-depth Analysis}
\label{In-depth Analysis}
\ 
\newline
\textbf{Generalization of DOGR.}
% 我们验证了DOGR的通用性，即对于常见的标识符构建方式，是否都能带来一定的提升。
We verify the generalization of DOGR, i.e. whether it can provide a certain improvement for common identifier construction methods including both lexical and numerical type in generative retrieval.
Table \ref{Different Identifier} presents the improvements of DOGR for common identifiers on NQ320k, along with a comparison to two other competitive methods.
We find that DOGR significantly improves the performance of lexical and numerical identifiers. 
For keyword DocID, we improve Recall@1 from 68.2 to 70.2 (a relative increase of 2.9\%). Meanwhile, for other DocID types, DOGR achieves at least a 2.1\%, 1.2\% and 1.7\% enhancement in Recall@1, Recall@10, and MMR@100 respectively, and demonstrates the generalization of the framework proposed. 
Additionally, DOGR outperforms two other methods that introduce additional objectives to capture query-document correlation in generative retrieval, achieving at least a 1.0\% improvement in R@1. 
This may be attributed to integrating the generation probability of identifiers with semantic representations, allowing for a more comprehensive modeling of the query-document relevance.
\begin{table}[!t]
\setlength{\tabcolsep}{1.2mm}{
\begin{tabular}{c|c|ccc}
\hline
 & \multicolumn{3}{c}{NQ320K(7830)} \\
\multicolumn{1}{c|}{DocID Type} & \multicolumn{1}{c|}{Enhance Type} & R@1 & R@10 & MRR@100 \\ 
\hline
\multirow{4}*{keyword} & \textbf{DOGR} & \textbf{70.2} & \textbf{89.1} & \textbf{76.8} \\
~ & LTRGR & 69.2 & 88.3 & 76.3 \\
~ & GLEN & 68.7 & 88.0 & 75.9 \\
~ & - & 68.2 & 87.7 & 75.5 \\ \hline
\multirow{4}*{$n$-gram} & \textbf{DOGR} & \textbf{67.2} & \textbf{86.9} & \textbf{74.3} \\
~ & LTRGR & 66.2 & 86.2 & 73.5 \\
~ & GLEN & 65.4 & 85.8 & 72.8 \\
~ & - & 65.2 & 85.4 & 72.8 \\ \hline
\multirow{4}*{first $n$-tokens} & \textbf{DOGR} & \textbf{66.3} & \textbf{86.7} & \textbf{73.9}  \\
~ & LTRGR & 65.4 & 85.7 & 73.0 \\
~ & GLEN & 64.5 & 85.1 & 72.3 \\
~ & - & 64.3 & 84.8 & 71.8 \\ \hline
\multirow{4}*{numerical} & \textbf{DOGR} & \textbf{67.8} & \textbf{86.8} & \textbf{74.9} \\
~ & LTRGR & 67.1 & 86.1 & 74.3 \\
~ & GLEN & 66.2 & 85.9 & 73.5 \\
~ & - & 66.4 & 85.7 & 73.6 \\
\hline
\end{tabular}
}
\caption{Retrieval performance of different methods on various identifier types. All the numbers are percentage numbers with `$\%$` omitted. Models with only generation task are denoted as `–'.}
\label{Different Identifier}
\end{table}
\begin{table}[!t]
\setlength{\tabcolsep}{0.9mm}{
\begin{tabular}{c|c|ccc}
\hline
 & &  \multicolumn{3}{c}{NQ320K(7830)} \\
\multicolumn{1}{c|}{Identifier length} & Conflict Rate & R@1 & R@10 & MRR@100 \\ 
% \multirow{4}{*}{Amazon Books} & MiceRec(K=2) & 10.384 & 16.344 & 20.781 \\
\hline
$l=3$ & 46.9 & 66.7 & 87.2 & 75.3 \\
$l=4$ & 28.6 & 67.9 & 88.0 & 75.7 \\
$l=6$ & 8.9 & 68.9 & 88.7 & 76.2 \\
$l=8$ & 3.0 & \textbf{70.2} & \textbf{89.1} & \textbf{76.8} \\
$l=10$ & 1.1 & 69.8 & 88.6 & 75.8 \\
$l=12$ & 0.4 & 69.6 & 88.3 & 75.5 \\ 
\hline
\end{tabular}
}
\caption{Impact of Identifier length on NQ320k. All the numbers are percentage numbers with `$\%$` omitted.}
\label{Impact of Identifier length}
\end{table}
\\
\textbf{Impact of Identifier length.} 
%Here we explore the impact of different identifier lengths on the retrieval performance. 
Since our framework allows different documents to share the same identifier, it may lead to identifier conflicts. Table \ref{Impact of Identifier length} shows the impact of identifier length on performance of NQ320k, where the conflict rate is defined as the ratio of the number of documents with identifier conflicts to the total number of documents. We find that when the identifier length is short, there are significant identifier conflict issues, which reduces the learning ability of the identifier generation task while increasing the ranking difficulty at the document level. Additionally, the shorter the identifier, the more limited its ability to convey information about the document. These two reasons may lead to a decline in model performance when the identifiers is short. Specifically, when the identifier length decreases from 8 to 3, Recall@1 drops by 5.0\%. 
On the other hand, as the length of the identifiers increases, the issue of identifier conflicts gradually diminishes, and the performance improves. Once the length reaches 8, the Recall@1 does not improve further.
One possible reason is that the model needs to memorize the order of keywords as the length increases, this may lead to a decrease in robustness.
% have little effect on capturing the document information. 
Moreover, the training and inference speeds also decrease with increasing identifier length.
\\
\textbf{Ablation Study.}
We also conducted detailed experiments to investigate the roles of different strategies in the training and inference phase: 
\begin{itemize}
    \item ``w/o doc segment sampling": The document segment sampling is removed in identifier generation stage.
    \item ``w/o prefix-ori contrastive / retrieval-aug contrastive / auxiliary generation loss": The prefix-oriented contrastive / retrieval-augmented contrastive / auxiliary generation loss is removed in document ranking stage, respectively.
    % \item "w/o generation loss": We removed the generation loss in document-level ranking stage.
    \item ``w/o semantic score": Semantic score is removed during the inference phase. This means that ranking is performed only based on generation probability, and in particular, random ranking is used when encountering identifier conflicts.
    \item ``w/o generation probability": The generation probability is removed during the inference phase. This means that after generating identifiers, the corresponding documents are ranked only based on the semantic score.
    \item ``w/o fusion rank": We adopt the same strategy as \cite{GLEN} instead of ranking by the fusion relevance score, i.e. overall ranking by generation probability and only rank by semantic score when encountering identifier conflicts.
\end{itemize}
\begin{table}[!t]
\setlength{\tabcolsep}{0.5mm}{
\begin{tabular}{c|ccc}
\hline
 & \multicolumn{3}{c}{NQ320K(7830)} \\
\multicolumn{1}{c|}{Model} & R@1 & R@10 & MRR@100 \\ 
% \multirow{4}{*}{Amazon Books} & MiceRec(K=2) & 10.384 & 16.344 & 20.781 \\
\hline
\textbf{DOGR} & \textbf{70.2} & \textbf{89.1} & \textbf{76.8} \\ \hline
w/o doc segment sampling & 68.6 & 88.2 & 75.9 \\ \hline
w/o prefix-ori contrastive loss & 69.0 & 88.4 & 76.0 \\
w/o retrieval-aug contrastive loss & 69.3 & 88.5 & 76.2 \\
w/o auxiliary generation loss & 65.8 & 84.3 & 72.9 \\ \hline
w/o semantic score & 68.2 & 87.7 & 75.5 \\
w/o generation probability & 63.3 & 87.6 & 72.0 \\
w/o fusion rank & 69.9 & 88.8 & 76.5 \\ \hline
\end{tabular}
}
\caption{Ablation study of \textbf{DOGR} with different strategy on NQ320k. All the numbers are percentage numbers with `$\%$` omitted.}
\label{Ablation Study}
\end{table}
The effectiveness of various strategies in \textbf{DOGR} on NQ320k is demonstrated in Table \ref{Ablation Study}, which provides the following insights:
\begin{itemize}
    \item Sampling of document segments can help the model capture the relationship between documents and identifiers more comprehensively, enhancing robustness and yielding a 2.3\% improvement in Recall@1. 
    \item Two contrastive learning tasks extract negative samples from different perspectives, resulting in gains of 1.7\% and 1.3\% in Recall@1, respectively. This indicates that contrastive learning can effectively enhance generative retrieval by document ranking. Additionally, we found that without the assistance of generative loss, the generative task experiences a decline, with Recall@1 dropping by 6.3\%, highlighting the necessity of generative loss.
    \item % Therefore, semantic information cannot be directly used as the basis for re-ranking. 
    During the inference phase, relying solely on the generation probability cannot address identifier collision issues, which results in a 2.8\% drop in Recall@1. On the other hand, discarding the generation probability in ranking causes a 9.8\% drop in Recall@1, which is mainly due to the lack of interaction between query and documents. Moreover, integrating both the generation probability and semantic information yields a further 0.4\% improvement in Recall@1 compared to ranking solely based on semantic information under identifier collisions.
\end{itemize}
\section{Conclusion}
In this paper, we propose \textbf{DOGR}, a novel and general framework that enhancing generative retrieval via document-oriented contrastive learning. 
It employs a two-stage learning strategy that simultaneously models the relevance between the query and documents by integrating the generation probability of identifiers with document semantic representation. 
% Considering the particularity of generative retrieval,  are implemented to enhance document representation. 
Moreover, two negative sampling methods and contrastive learning objectives are employed to comprehensively learn the relationship between the query and the document through their direct interaction.
% Furthermore, 相应的对比学习目标 通过query和document的直接交互来学习两者的相关性，
% Considering the particularity of generative retrieval, contrastive learning from
% different views are introduced to optimize the
% learning gap between the generation and ranking objective through exploring hard negatives, as well as to tackle the document ranking problem when encountering identifier conflict.
Experimental results demonstrate that \textbf{DOGR} achieves state-of-the-art among generative retrieval methods on two benchmark datasets and brings improvements on common identifier construction methods.
In the future, we will continue to explore how to end-to-end integrate information from documents to enhance the retrieval capabilities of generative retrieval in large-scale corpora.

\bibliography{aaai25}

\end{document}